# Laser scanning reflection-matrix microscopy for label-free *in vivo* imaging of a mouse brain through an intact skull

Seokchan Yoon[1,2,+], Hojun Lee[1,2,+], Jin Hee Hong[1,2], Yong-Sik Lim[3] and Wonshik Choi[1,2,*]

[1]*Center for Molecular Spectroscopy and Dynamics, Institute for Basic Science, Seoul 02841, Korea*
[2]*Department of Physics, Korea University, Seoul 02855, Korea*
[3]*Department of Nano Science and Mechanical Engineering and Nanotechnology Research Center, Konkuk University, Chungbuk, Korea.*
[*]wonshik@korea.ac.kr

**Abstract**
We present a laser scanning reflection-matrix microscopy combining the scanning of laser focus and the wide-field mapping of the electric field of the backscattered waves for eliminating higher-order aberrations even in the presence of strong multiple light scattering noise. Unlike conventional confocal laser scanning microscopy, we record the amplitude and phase maps of reflected waves from the sample not only at the confocal pinhole, but also at other non-confocal points. These additional measurements lead us to constructing a time-resolved reflection matrix, with which the sample-induced aberrations for the illumination and detection pathways are separately identified and corrected. We realized in vivo reflectance imaging of myelinated axons through an intact skull of a living mouse with the spatial resolution close to the ideal diffraction limit. Furthermore, we demonstrated near-diffraction-limited multiphoton imaging through an intact skull by physically correcting the aberrations identified from the reflection matrix. The proposed method is expected to extend the range of applications, where the knowledge of the detailed microscopic information deep within biological tissues is critical.

**Introduction**
Confocal detection is one of the most-widely used configurations in deep-tissue optical microscopy due to its ability to reject signals from out-of-focus planes and other unwanted noise due to multiple light scattering[1,2]. It has been employed either by itself in confocal reflectance/fluorescence imaging[3] or combined with the temporal gating to further attenuate multiple scattering noise[4]. However, in the presence of the sample-induced aberrations, signals containing the object information get spread away from the confocal pinhole due to the blur of the point-spread-function (PSF)[5–8]. Image contrast is reduced as a consequence, and the resolving power is gradually lost with the increase of imaging depth. The key to resolving the sample-induced aberrations and achieving the ideal diffraction-limited imaging deep within scattering medium is to coherently refocus the non-confocal signals, i.e. signals arriving at positions other than the confocal pinhole, back to the confocal detection position. Since multiple scattering noise arrives at the non-confocal positions as well as the spatially aberrated signals, it is critical to selectively refocus the aberrated signals, not the multiple scattering noise. However, this has been extremely difficult in the optical coherence imaging, when compared with fluorescence imaging, mainly because the signal and multiple scattering noise have identical frequencies.

The most straightforward approach for focusing non-confocal signals back to the confocal position is to place a wavefront shaping device such as a deformable mirror or a liquid-crystal spatial light modulator (SLM) in the illumination and/or detection beam paths. For the selective focusing of the blurred signal, the SLM adjusts the wavefront iteratively in such a way to maximize the intensity or sharpness of the resulting confocal image[9–13]. Although these so-called sensorless adaptive optics (AO) methods are capable of finding high-order aberrations, mainly in the fluorescence imaging, the time taken for the feedback process tends to be lengthy as the image acquisition is required for each iteration step. A model-based modal aberration correction approach may reduce the time for the optimization, but it works for only a few lowest modes of Zernike polynomials[14]. Furthermore, the iterative optimization can only be initiated when the objects are visible in the first place.

To correct aberrations with the minimal number of image acquisitions, various approaches have been proposed based on the direct measurement of the wavefront of the backscattered signals, which are referred to as wavefront-sensing AO. Shark-Hartmann wavefront sensor[15–19] and interference microscopy[20–23] have been widely used for the wavefront recording. In general, the sample-induced aberrations for both the illumination and detection beam paths contribute to the measured wavefront. Mathematically, the detection PSF is convoluted with the multiplication of an object function and illumination PSF, which makes it difficult to extract the individual PSFs. Most previous studies resolved this double-pass problem in the limiting case when there exists a bright point-like scatterer within the sample, similar to the guide stars in the astronomy[6]. This makes the multiplication of the object function and illumination PSF a point source regardless of the illumination PSF, and the measured wavefront aberration becomes a measure of the single-pass aberration in the detection beam path. Guide stars are not always available, but photo-receptor cells in retina imaging[18] and nonlinear excitation can serve as intrinsic and approximate guide stars[15,16,19,24,25]. A strategy working in another limiting case is to illuminate a planar wave, not the focused wave, for minimizing aberrations in the illumination beam path. Then, aberrations can be found from a reflected wavefront by computationally optimizing the image metrics such as image sharpness[22,23,26]. The drawback of this approach is its susceptibility to the multiple scattering noise due to the loss of confocal gating. The main reason why the existing wavefront-sensing AO approaches work in the limiting cases is due to the incomplete recording of the input-output response of the specimens of interest, which makes the double-pass problem underdetermined.

Here, we propose a laser scanning reflection-matrix microscopy (LS-RMM) that records non-confocal signals as well as confocal signals in their phases and amplitudes. These measurements constitute a reflection matrix quantifying the input-output response of the light-medium interaction. Specifically, the proposed system scans a focused illumination to the sample and records the wide-field map of the reflected waves in the full image plane, not just at the single confocal position. An off-axis low-coherence interferometry is employed for the time-gated detection of the backscattered waves, which excludes multiple scattering noise arriving at different flight times from the signal waves. Conventional optical coherence microscopy (OCM) that combines confocal microscopy and optical coherence tomography is incapable of resolving the round-trip aberrations as the problem is underdetermined due to the incomplete data acquisition. On the other hand, LS-RMM takes advantage of the non-confocal signals to unambiguously identify the highly complex aberrations from the roundtrip distortions and yet maintain the confocal gating capability. The key step is to transform the measured reflection matrix taken for the focus illuminations to that for the planar illuminations by the linear superposition of complex-field maps taken for the focused illuminations. We then applied the unique algorithm termed closed-loop accumulation of single scattering (CLASS) reported earlier to selectively identify and computational correct the sample-induced aberrations from the background multiple scattering noise[27–29].

The LS-RMM proposed in this study is different from the previous reflection matrix based approaches in its illumination and detection configurations[27,28,30]. Our earlier studies used planar wave illuminations, while the present LS-RMM uses a focused illumination configuration. The main benefit is the enhanced signal to multiple-scattering noise ratio. In the previous planar wave illumination, multiple scattering noise generated by the illumination over wide area is all added together at each detection pixel. On the other hand, signals generated by a focused illumination compete only with the multiple scattering noise generated by the given focused illumination itself[31]. Other approaches employing focused illumination usually place a detector in the pupil plane where the signal are spread over the detection area[20,32]. On the contrary, we placed a wavefront detector at the image plane where the signals are more likely to be focused relative to the multiple scattering noise, which gives the best spatial distinction between the signal and multiple scattering noise and also gives an additional advantage in making efficient use of detector dynamic range. Due to the enhanced sensitivity with respect to the multiple scattering noise, we could achieve in vivo imaging of cortical myelination in the mouse brain through the intact skull, an extreme form of a complex scattering medium.

Another important benefit of the focused illumination geometry lies in its direct compatibility with the existing imaging modalities such as OCM and multi-photon microscopy (MPM). Our AO system was integrated with these conventional imaging modalities simply by inserting a wide-field and low-coherence interferometry in the detection plane. Furthermore, it also combines hardware adaptive optics

(HAO) and computational adaptive optics (CAO)[21] together to complement each other. We demonstrated hardware correction of aberrations that are computationally identified by CLASS algorithm by using a SLM in the beam path. We realized near diffraction-limited second-harmonic-generation (SHG) imaging of collagen fibers through an intact mouse skull. We would like to emphasize that the realization of LS-RMM in the focused illumination configuration has been technically challenging due to shot-to-shot phase fluctuations between the images for focused illuminations, especially for in vivo imaging applications. We maintained phase-referencing between multiple illuminations by improving mechanical stability of the interferometry and laser scanning system.

**Results**
**Experimental schematic of the laser scanning reflection-matrix microscopy (LS-RMM)**
The LS-RMM is based on the OCM system, but a camera was placed in a plane conjugate to the image plane instead of a confocal pinhole and a photodetector. A reference plane wave was introduced to the camera to cover a full detection area, and the time-gated electric-field (E-field) images were recorded for waves backscattered from the sample by means of off-axis low-coherence interferometry. A schematic diagram of the imaging system is shown in Fig. 1a. A mode-locked Ti:Sapphire laser (center wavelength: 900 nm, bandwidth: 25 nm, and repetition rate: 80 MHz) was used as a short-coherence-length light source (coherence length: 30 μm). The laser beam was split into the sample and reference waves by a beamsplitter (BS1), and they were recombined by another beamsplitter (BS2) to form a Mach-Zehnder interferometry. The sample beam was relayed via two galvanometer-scanning mirrors (GMs) and an SLM (X10568-02, Hamamatsu) and focused on the sample plane by a water immersion objective lens (OL, Nikon, x60, NA 1.0). The GMs were used to raster-scan the focused illumination beam at the sample as in conventional confocal microscope, and the SLM to physically correct aberration maps retrieved by CLASS algorithm. They were all located at planes conjugate to the back pupil of OL. The backscattered wave from the sample was collected by the same OL and traveled through the reciprocal path. It was reflected again by the SLM, de-scanned by the GMs, and finally delivered to the camera (s-CMOS, pco.edge 4.2, PCO AG) located at a conjugate image plane. In the case of the reference wave, it was sent through multiple pairs of relay lenses, an optical delay line (not shown) and a diffraction grating (DG). The first-order diffraction by the DG was selected by an iris diaphragm (I) and delivered to the camera as a plane wave. Consequently, its phase front is tilted with respect to the camera plane while its temporal pulse front remained parallel to the camera plane. The interference between the sample and reference waves form an off-axis interferogram, from which we could retrieve amplitude and phase images of the sample waves at the temporal gating width corresponding to 15 μm, half the coherence length of the light source. For the image acquisition, the focused illumination beam was scanned in a raster mode by the two GMs over a given field of illumination (FOI) area on the sample plane with a scanning interval of $\lambda/2\alpha$, where $\alpha$ is the numerical aperture of the OL. This FOI is equivalent to the field of view of the conventional OCM.

For each illumination point $\mathbf{r}_{in}$, a time-gated E-field image of the backscattered wave $E_{cam}(\mathbf{r}_{cam}; \mathbf{r}_{in})$ was obtained at the camera plane $\mathbf{r}_{cam}$ over a certain detection area termed as a field of detection (FOD). For example, Figs. 1b and 1c show the intensity and phase images, respectively, taken in the absence of aberration. The width of the focus was about 450 nm, equal to the spread of the diffraction-limited PSF. The phase is meaningful only near the focused spot. Figure 1d shows the intensity map taken in the presence of an aberrating medium, showing significantly broadened distribution with the peak intensity attenuated by more than an order of magnitude. The corresponding phase map (Fig. 1e) also showed a broadened distribution of meaningful phase values. In contrast to previous wavefront sensing AO where a wavefront detector was located at a conjugate pupil plane[20], the present scheme offers the best spatial distinction between the signal and multiple scattering noise unless the aberration is too severe to broaden the PSF over the entire FOD. Here, FOD was adjusted to be wide enough to capture the entire profiles of the reflected wave blurred by the sample-induced aberration. The choice of the optimal sizes of FOD and FOI requires the consideration of the PSF broadening, the intensity ratio between single- and multiple-scattered waves, and image acquisition time. The wider FOD are used, higher spatial frequency aberration can be measured, but more multiple scattering noise is captured. In the experiment, they were appropriately adjusted depending on the properties of the aberration and scattering of the sample. The image acquisition time is mainly determined by a frame rate of the camera. For a FOD of 50×50 μm², which gives ~10,000 effective number of correction modes in the pupil, the total acquisition time for a full reflection matrix for a FOI of 50×50 μm² is about 15 seconds (see Methods in details). The SLM serves as a flat mirror for the reflection matrix recording and the image acquisition of label-free reflectance imaging. It was actively used for SHG imaging by physically compensating the sample-induced aberrations identified by CLASS algorithm.

**Image processing and aberration correction**

In the case of optical coherence imaging, the sample-induced aberrations of the incident wave on the way to the sample and those of the reflected wave on its way back to the detector have to be corrected separately. This necessitates the recording of a reflection matrix and the application of CLASS algorithm. In this section, we described how to process a sequence of images taken by LS-RMM for constructing a time-gated reflection matrix in the space domain. For a given focused illumination at a point $\mathbf{r}_{in}$ at the sample plane, the time-gated complex-field map of the reflected wave $E_{cam}(\mathbf{r}_{cam}; \mathbf{r}_{in})$ recorded in the camera plane $\mathbf{r}_{cam}$ takes the form of

$$E_{cam}(\mathbf{r}_{cam}; \mathbf{r}_{in}) = \int Q(\mathbf{r}_{cam}; \mathbf{r}) \cdot [O(\mathbf{r}) \cdot P(\mathbf{r}; \mathbf{r}_{in})] d^2\mathbf{r} + E_m(\mathbf{r}_{cam}; \mathbf{r}_{in}). \qquad (1)$$

Here $P(\mathbf{r}; \mathbf{r}_{in})$ is the E-field PSF at a point $\mathbf{r}$ in the sample plane, which is spread with respect to $\mathbf{r}_{in}$ due to the sample-induced aberrations. $O(\mathbf{r})$ is the object function given by the amplitude reflectance of the target object. $Q(\mathbf{r}_{cam}; \mathbf{r})$ describes the E-field PSF of the returning waves from the sample plane to the camera plane. $P$ and $Q$ are theoretically the same in the epi-detection geometry, but they can be different due to the slight mismatch of optical system between the illumination and detection optics. $E_m(\mathbf{r}_{cam}; \mathbf{r}_{in})$ represents a speckle field due to the multiple-scattered waves by the scattering layer above the sample plane.

To identify the sample-induced aberrations, we need to obtain the complex-field map of the reflected wave $E_{lab}(\mathbf{r}_o; \mathbf{r}_{in})$ in the laboratory frame coordinate $\mathbf{r}_o$. Since $E_{cam}(\mathbf{r}_{cam}; \mathbf{r}_{in})$ is acquired after the de-scanning action by the GMs, $\mathbf{r}_o$ and $\mathbf{r}_{cam}$ have the relation, $\mathbf{r}_o = \mathbf{r}_{cam} + \mathbf{r}_{in}$. Therefore, E-field in the laboratory frame can be obtained from $E_{lab}(\mathbf{r}_o; \mathbf{r}_{in}) = E_{cam}(\mathbf{r}_o - \mathbf{r}_{in}; \mathbf{r}_{in})$. Figure 2b shows a set of $E_{lab}(\mathbf{r}_o; \mathbf{r}_{in})$ at the laboratory frame for the scanning of illumination spots indicated by the white × marks. The data was taken for a custom-made resolution target placed under a 600-μm-thick rough-surfaced plastic layer exhibiting strong aberrations (Fig. 2a). With the set of images in Fig. 2b, we constructed a time-gated reflection matrix $\mathbf{R}(\mathbf{r}_o; \mathbf{r}_{in})$ in the space domain by assigning $E_{lab}(\mathbf{r}_o; \mathbf{r}_{in})$ to each column of $\mathbf{R}$. Therefore, the column and row indices of $\mathbf{R}$ are $\mathbf{r}_{in}$ and $\mathbf{r}_o$, respectively (Fig. 2c). The acquired reflection matrix $\mathbf{R}$ can be represented as

$$\mathbf{R} = \mathbf{QOP} + \mathbf{M}, \qquad (2)$$

using the transmission matrix formalism. Here $\mathbf{O}$ is a diagonal matrix containing the target's object function in the diagonal elements, i.e. $\mathbf{O}(\mathbf{r}; \mathbf{r}) = O(\mathbf{r})$. $\mathbf{P}$ is the input PSF matrix whose element is given by $P(\mathbf{r}; \mathbf{r}_{in})$, and the output PSF matrix $\mathbf{Q}$ consists of $Q(\mathbf{r}_o; \mathbf{r})$. $\mathbf{M}$ is the multiple scattering matrix composed of $E_m(\mathbf{r}_o; \mathbf{r}_{in})$. The conventional confocal image can be obtained from the main diagonal elements of $\mathbf{R}$ (Fig. 2f). A strong spatial gating rules out the multiple scattering noise because it takes only one pixel corresponding to the confocal pinhole. On the other hand, the confocal image was blurred and severely distorted due to the pronounced sample-induced aberrations contained in $\mathbf{P}$ and $\mathbf{Q}$. In order to reconstruct an aberration-free object image, we need to identify and remove $\mathbf{P}$ and $\mathbf{Q}$ in $\mathbf{R}$.

As the sample-induced aberrations are mainly dominated by the angle-dependent phase retardation, we converted the position-basis of the reflection matrix $\mathbf{R}$ into the spatial frequency basis by applying a unitary Fourier transform operator $\mathbf{F}$ to both input and output sides of $\mathbf{R}$, i.e. $\widetilde{\mathbf{R}} = \mathbf{F}^{-1}\mathbf{RF}$. The reflection matrix $\widetilde{\mathbf{R}}$ thus obtained in the spatial frequency basis (k-space) is shown in Fig. 2d. Mathematically, this can be expressed as

$$\widetilde{\mathbf{R}} = \widetilde{\mathbf{Q}}\widetilde{\mathbf{O}}\widetilde{\mathbf{P}} + \widetilde{\mathbf{M}}, \tag{3}$$

were $\widetilde{\mathbf{O}} = \mathbf{F}^{-1}\mathbf{OF}$ is a circulant matrix whose element $\widetilde{O}(\mathbf{k_o}; \mathbf{k_{in}}) = O(\mathbf{k_o} - \mathbf{k_{in}})$. $\widetilde{\mathbf{P}} = \mathbf{F}^{-1}\mathbf{PF}$ and $\widetilde{\mathbf{Q}} = \mathbf{F}^{-1}\mathbf{QF}$ are the transmission matrices for a plane wave propagating through the illumination and detection pathways, respectively. For the area where the PSF is shift-invariant, i.e. within isoplanatic patch, the relation $P(\mathbf{r}; \mathbf{r_{in}}) = P(\mathbf{r} - \mathbf{r_{in}})$ is valid. Then, $\mathbf{P}$ takes the form of a Toeplitz matrix, which sets $\widetilde{\mathbf{P}}$ a diagonal matrix whose diagonal element consists of $e^{-i\phi_{in}(\mathbf{k_{in}})}$. Likewise, $\widetilde{\mathbf{Q}}$ is a diagonal matrix with diagonal element given by $e^{-i\phi_o(\mathbf{k_o})}$. Here $\phi_{in}(\mathbf{k_{in}})$ and $\phi_o(\mathbf{k_o})$ are the angle-dependent phase retardations induced by the scattering layer for the illumination and detection pathways, respectively.

We applied CLASS algorithm to $\widetilde{\mathbf{R}}$, which separately identifies $\phi_{in}(\mathbf{k_{in}})$ and $\phi_o(\mathbf{k_o})$ in such a way that the total intensity of the confocal image reconstructed from an aberration-corrected reflection matrix[27,28]. The retrieved $\phi_{in}(\mathbf{k_{in}})$ and $\phi_o(\mathbf{k_o})$ are shown in Figs. 2h and 2i, respectively. Conjugations of the pupil functions were applied to the original matrix $\widetilde{\mathbf{R}}$ in Fig. 2d to obtain the aberration-corrected matrix $\widetilde{\mathbf{R}}^{(c)} = \widetilde{\mathbf{Q}}^*\widetilde{\mathbf{R}}\widetilde{\mathbf{P}}^*$, which was then transformed back to the position-space matrix $\mathbf{R}^{(c)}$ (Fig. 2e). Due to the compensation of the sample-induced aberrations, the magnitude of the diagonal elements of $\mathbf{R}^{(c)}$ was greatly enhanced and that of the off-diagonal elements was reduced in comparison to the original matrix $\mathbf{R}$ shown in Fig. 2c. The image acquired from the diagonal elements of $\mathbf{R}^{(c)}$ is shown in Fig. 2g, where image intensity was increased by about 30 times in comparison with the original confocal image in Fig. 2f. And a spatial resolution of the corrected image was estimated to be 450 nm, identical to the ideal diffraction limit.

**Physical aberration correction by using SLM**
The LS-RMM system was designed not only for identifying the high-order aberrations induced by the medium, but also for physically compensating them to generate a near-diffraction-limited focus within the sample. To this end, the SLM in the experimental setup (Fig. 1a) was used to display the phase conjugation of the identified aberration map shown in either Fig. 2h or 2i. As illustrated in Fig. 2j, the SLM compensates wavefront distortions induced by the sample-induced aberrations for the wave incident to the sample. The reflected wave from the sample is corrected as well by the same SLM to generate a sharp focus at a detector plane. The original PSF of the reflected wave obtained from the OCM image in Fig. 2b was highly speckled and broadly spread in width. After the hardware wavefront correction by the SLM, the PSF became a sharp focus with a significantly increased peak intensity (Fig. 2k). To assess the quality of the aberration correction, we measured the intensity profile across the center of the focus (red curve in Fig. 2l). The enhancement of Strehl ratio, measured by the ratio of the peak intensities after and before aberration correction, was 18, and the width of the focus was almost close to the diffraction-limited spatial resolution of 450 nm. The peak height reached about 20 % of the computational aberration correction by CLASS algorithm (blue curve in Fig. 2l). This is mainly due to the limitation of the SLM in shaping steeply varying aberration especially in the high spatial frequency range. However, the width of PSF was almost identical to that of the computational correction. With hardware wavefront correction by the SLM in place, we could obtain a clean OCM image (Fig. 2m) with no need of computational correction by the CLASS algorithm. The quality of image is comparable to that seen in the CLASS image (Fig. 2g), which confirms that a sharp focus was physically generated at the plane of the target object.

**In vivo imaging of mouse brain through an intact skull**
With the capability of LS-RMM to deal with the extreme high-order aberrations, we attempted to image neuronal structures in the mouse brain through the intact skull. The mouse skull contains full of fine microstructures, thereby inducing severe optical aberrations as well as strong multiple scattering noise. So far, only multi-photon fluorescence imaging could see through a mouse skull because only the input wavefront aberration matters in multi-photon microscopy and nonlinear excitation helps to reduce the image blur[10,24,33,34]. On the contrary, label-free reflectance imaging has not yet been able to image below an intact skull especially because the roundtrip aberrations jointly deteriorate the image quality in the reflectance imaging. Furthermore, the mouse skull consists of several layers of microstructures, which reduces the size of the isoplanatic patch down to the degree that conventional AO methods can hardly keep up with[35]. In our analysis, we chose the region of analysis area adaptively depending on the isoplanatic patch size and applied CLASS algorithm to obtain local position-dependent aberrations.

Figure 3 shows the reflectance images of myelinated axons in the cortex layer 1 for various types of sample preparation. Initially, we performed LS-RMM imaging through a thinned skull (Figs. 3a-d). Compact bone and spongy bone were removed from the skull of a 7-week-old mouse to reduce the thickness of the skull to approximately 40 μm (Fig. 3a). We recorded a time-gated reflection matrix **R** with a FOD of 50×50 μm$^2$ for a FOI of 50×50 μm$^2$ at the depth of 72 μm from the upper surface of the skull. In a conventional OCM image (Fig. 3b) reconstructed by diagonal elements of **R**, the shape of myelinated axons appeared to be blurred. We divided entire view field into 6×6 regions and corrected aberrations for each region. The patch size of each region was approximately 10×10 μm$^2$ including margins overlapping with neighboring regions. The number of correction modes in each aberration map was about 10,000. Myelinated axons were visible clearly with the improved sharpness and signal strength in the aberration corrected image shown in Fig. 3c. The aberrations varied spatially, but the degree of aberration was rather mild as seen in the enlarged aberration map due to the thinning of the skull.

As a next step, we performed ex vivo imaging of a 3-weeks-old mouse with its skull intact. In the 3D image reconstructed by SHG signals detected at the PMT, a compact bone, spongy bone and meninge could be identified (Fig. 3e). Thickness of the skull was about 80-100 μm. After measuring **R** at the depth indicated by the red dotted box, approximately 125 μm below from the upper surface of the skull, the OCM image (Fig. 3f) and LS-RMM image (Fig. 3f) were obtained along with local aberrations (Fig. 3h). Analysis with the same patch size as thinned skulls yields high resolution, high contrast image in the aberration corrected images. Myelin thickness was measured to be as small as 450 nm. Because the intact skull was thicker than the thinned skull and composed of multiple layers, the complexity of aberrations was increased and the correlation between the aberration maps of neighboring regions was reduced.

We demonstrated the imaging of myelinated axons through an intact skull of a living mouse (Fig. 3i-l). After removing the scalp of an 8-week-old mouse as shown in Fig.3i, a 100-μm-thick cover glass was glued on the surface of the intact mouse skull. A skull holder was attached to the mouse head with dental cement, which was then firmly fixed onto the custom-built stage (see Methods in details). The thickness of the skull was measured to be approximately 120 μm. At 200 μm below the skull surface, the OCM image was so noisy that no structures were visible at all (Fig. 3j). As shown in Fig. 3k, myelinated axons were made visible after the aberration correction in each region divided by 10×10 μm$^2$ patch size. The aberration maps in each patch area (Fig.3l) were so complex that the number of effective orthogonal modes required to represent such aberration maps amounted to 10,000. Furthermore, the aberration maps between neighboring areas were completely decorrelated. It is worth noting that the application of CLASS algorithm to the patch size greater than 10×10 μm$^2$ failed to resolve any structures, confirming the degree of local aberrations.

**Wavefront shaping for nonlinear imaging through an intact skull**
As the LS-RMM can physically generated a sharp focus deep within complex scattering and aberrating media, it can be directly combined with multi-photon microscopy such as two-photon microscopy and second-harmonic-generation (SHG) microscopy. The experimental setup needs little modification, and all needed is to place a PMT right behind the dichroic mirror (Fig. 1a). Here, we demonstrated near-diffraction-limited SHG imaging through an intact skull. We placed collagen gel matrix under a 100-μm-thick mouse skull extracted from a 3-week-old mouse (Fig. 4a). The objective focus was set 240 μm below the bottom surface of the skull. Due to complex aberrations by the skull, conventional OCM failed to visualize any collagen fibers (Fig. 4b). The collagen fiber structures were made clearly visible when CLASS algorithm was applied locally to 8×8-μm$^2$ sized region, which was the effective size of the isoplanatic patch of the skull (Fig. 4c). The number of correction modes in each aberration map was about 13,000. Figure 4d shows the aberration maps for each of 6×6 regions. The fine collagen fibrils were clearly resolved over the entire field of view (Fig. 4c).

Similar to the demonstration in Figs. 2j-m, we could physically correct the skull-induced aberrations by the SLM. Since the aberrations varied depending on the position, we made a physical correction for each subregion at a time. For example, we chose the aberration map indicated by the red circle in Fig. 4d, and its phase conjugation was written on the SLM. Figure 4e shows a conventional OCM image taken over the same area as that in Fig. 4b after correcting the local aberration. Only the collagen fiber structures associated with the area indicated by the red circle were resolved. The other areas didn't show any structures, supporting that the isoplanatic size was as small as $8\times8$ $\mu m^2$. Simultaneously, we took SHG images by using the PMT (Fig. 4g), which shows the same structures as those resolved by aberration-corrected OCM image, while no fiber structures were visible at all in the uncorrected SHG image as shown in Fig. 4f. The measured width of collagen fiber was as small as 500 nm, close to the diffraction limit.

**Conclusion**
We developed a laser scanning reflection-matrix microscopy that is readily combined with the conventional laser scanning confocal microscopy, and yet capable of correcting extremely high-order aberrations even in the presence of strong multiple scattering noise. While the focused illumination is raster-scanned at the sample plane as is done in confocal microscopy, the E-field of backscattered waves were detected over aa wide area instead of confocal detection through a pinhole. This allows us to use the unique algorithm based on the reflection matrix, with which we corrected sample-induced aberrations for more than 10,000 angular modes with no need of any guide stars. We demonstrated label-free reflectance imaging of cortical myelination through an intact skull of a living mouse with spatial resolving power close to the ideal diffraction limit. Furthermore, we physically corrected the aberrations identified from the reflection matrix by using SLM and demonstrated near-diffraction-limited SHG imaging of collagen fibers underneath a mouse skull.

In LS-RMM, the identification of wavefront aberrations is based on reflectance imaging relying on the intrinsic reflectance contrast of targets. Therefore, it does not require fluorescent labeling and high excitation power as opposed to the most existing AO modalities relying on multi-photon fluorescence feedback signals. Another critical advantage of LS-RMM is that the computational process to retrieve both the wavefront aberrations and aberration-corrected reflectance image takes place after the recording a reflection matrix. This decouples the measurement speed from the speed of the hardware wavefront shaper since it is used only one time for displaying the final correction map retrieved by an aberration correction algorithm. The most time-consuming part is the post computational process, which mainly depends on computing power, and this can be drastically improved by using GPU or cluster-based hardware and also by an improvement of the algorithm.

## Methods

**Acquisition time for a full reflection matrix:** A frame rate $R_f$ of the sCMOS camera for a given FOD scales inversely and linearly with a number of lines in the FOD; $R_f = r/\sqrt{\text{FOD}}$ [Hz], where the coefficient $r$ was 40,000 [1/μm] in the experiment. The number of effective correction modes (pixels) $N_c$ in a circular pupil with a numerical aperture $\alpha$ is given by $N_c = \pi(\sqrt{\text{FOD}}/2\delta_d)^2$, where $\delta_d = \lambda/2\alpha$ is the detection resolution. For example, the FOD size of 50×50 μm² contains an effective number of correction modes of about 10,000. The number of images to be recorded for the complete measurement of a reflection matrix for a given FOI is given by $N_s = \text{FOD}/\delta_d^2$. The total acquisition time for a full reflection matrix is then determined by $T = N_s/R_f$. For the FOI size of 50×50 μm² with a FOD of 50×50 μm² and $\alpha = 1$, the total acquisition time is about 15 seconds. With a reduced FOD size of 16×16 μm², which contains about 1,000 number of correction modes, the total acquisition time becomes about 5 seconds. The use of a high-speed CMOS camera can reduce the matrix acquisition time even below 1 seconds.

**Sample preparation for ex vivo imaging:** All animal experiments were approved by the Korea University Institutional Animal Care & Use Committee (KUIACUC-2019-0024). Three- or eight-week-old C57BL/6 mice were deeply anesthetized with an intraperitoneal injection of ketamine/xylazine (100/10 mg/kg) and decapitated. The scalp was removed, and the whole head was fixed with 4 % paraformaldehyde at 4 °C for 1~2 days and washed with PBS three times. For imaging through thinned skull window, the compact and spongy bone were ground from a fixed head of a 7-week-old mouse. The thinned skull window (2~3 mm in diameter) was made in the center of the parietal bone.

**Preparation of intact skull window for in vivo imaging:** C57BL/6 mice (7~9 weeks old, 22~28 g) were temperature controlled and anesthetized with isofluorane (1.5~2 % in oxygen for surgery to maintain a breathing frequency around 1 Hz). A dexamethasone (0.2 mg/kg body weight) was administrated intramuscularly to minimize swelling at the site of surgery and on the two consecutive days after the surgery. Eyes were covered with eye ointment during surgery and imaging. The scalp was removed to expose both parietal plates and the bregma and lambda. A sterile round coverslip of 4-mm diameter (#1 thickness) was attached to the center of parietal bone of skull using ultraviolet-curable glue (Loctite 4305). A custom-made metal plate was fixed to the skull for head fixation, and the exposed part of the skull was covered with dental cement.

**Preparation of collagen gel under mouse skull:** Collagen gel was made from rat tail collagen (Collagen type I, Corning, New York, USA). Collagen type I gel solution (2 mg/ml) was neutralized with 0.1 N NaOH and chilled into ice. To form the collagen gel, mixed solution was incubated at 37 °C for 60 min. Skulls were excised from 3-week-old C57BL/6 mice and quickly immerged in PBS. The skulls fixed with 4 % paraformaldehyde were washed with PBS and mounted on the collagen gel for imaging.

# Figures

## Figure 1

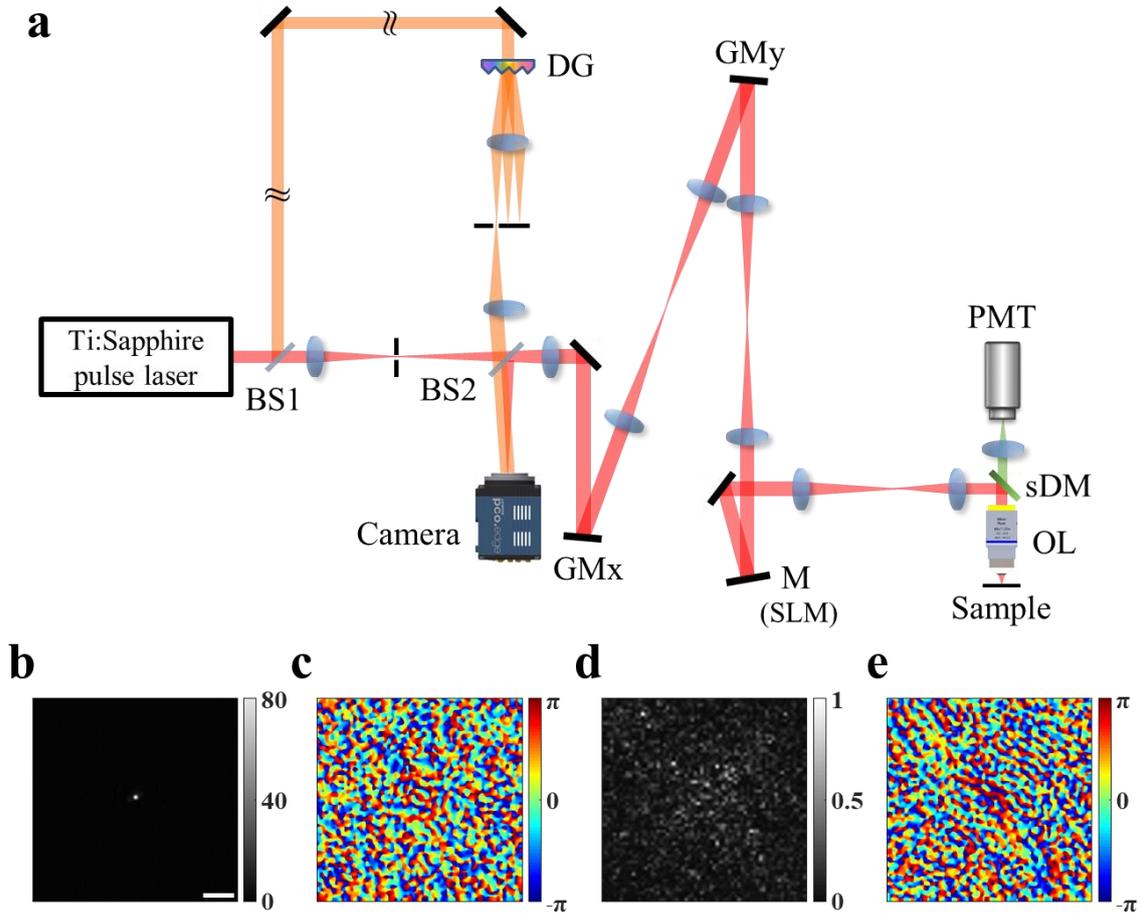

**Figure 1.** Experimental layout of the laser scanning reflection-matrix microscopy. **a,** Experimental schematic. BS1 and BS2: beam splitters. The sample and reference beams are colored in red and orange, respectively, for visibility. GMx and GMy: galvanometer mirrors for scanning the sample beam along *x* and *y* axes, respectively M: mirror, SLM: spatial light modulator, sDM: short-pass dichroic mirror, OL: objective lens, PMT: photomultiplier tube, DG: diffraction grating. P: pinhole. **b** and **c,** Intensity and phase images of reflected wave from the sample, respectively, measured without aberrations. **d** and **e,** Same as **b** and **c,** but for an aberrating sample. Color bars in **b** and **d** indicate intensity normalized by the maximum intensity in **d**. Color bars in **c** and **e,** phase in radians. Scale bar, 5 μm.

Figure 2

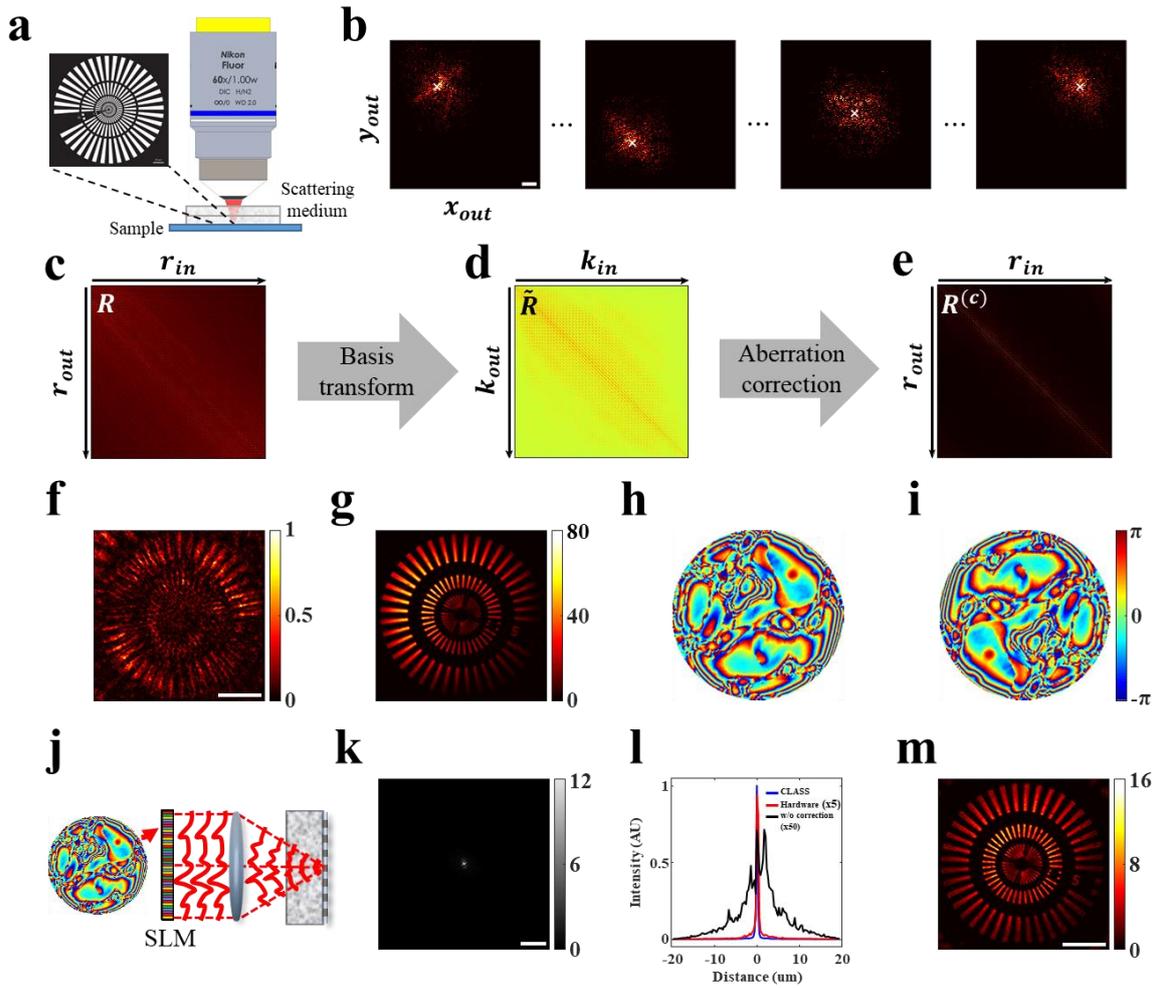

**Figure 2. Aberration correction by exploiting the time-resolved reflection matrix a,** Sample geometry. As a test sample, a homemade Siemens star target was placed under a 600-μm-thick rough-surfaced plastic layer exhibiting strong aberrations. **b,** A set of E-field images of reflected waves in the laboratory coordinate $\mathbf{r_o} = (x_o, y_o)$. Four representative amplitude images are shown here. The white × marks indicate illumination points. Scale bar, 10 μm. **c,** Time-resolved reflection matrix **R** in position space constructed from the set of E-field images in **b**. Each image was converted to a column vector and assigned to the corresponding column of **R**. **d,** Reflection matrix $\tilde{\mathbf{R}}$ in k-space **e,** Aberration-corrected reflection matrix $\mathbf{R}^{(c)}$ converted from $\tilde{\mathbf{R}}$ after the application of CLASS algorithm. **f,** OCM image constructed from the main diagonal of **R** in **c**, i.e. before aberration correction. **g,** Aberration-corrected CLASS image obtained from the main diagonal of $\mathbf{R}^{(c)}$ in **e**. Scale bar, 10 μm. **h** and **i,** Aberration maps of illumination and detection pupil functions retrieved by CLASS algorithm, respectively. The radius of the maps corresponds to $k_0\alpha$, with $\alpha = 1.0$ NA. The illumination and detection aberrations are almost identical as a consequence of the optical reciprocity principle. **j,** Schematic of physical aberration correction. The conjugate of the illumination pupil phase map in **h** was displayed on the SLM in Fig. 1a to physically compensate the aberrations. **k,** Intensity image of a reflected PSF measured at the camera after physically correcting the aberration by the SLM. Scale bar, 10 μm. **l,** Line profiles of the reflected PSFs obtained without wavefront correction (black), after computational wavefront correction by CLASS algorithm (blue), and after hardware wavefront correction by the SLM (red). **m,** The OCM image obtained after hardware wavefront correction. Scale bar, 10 μm.

Figure 3

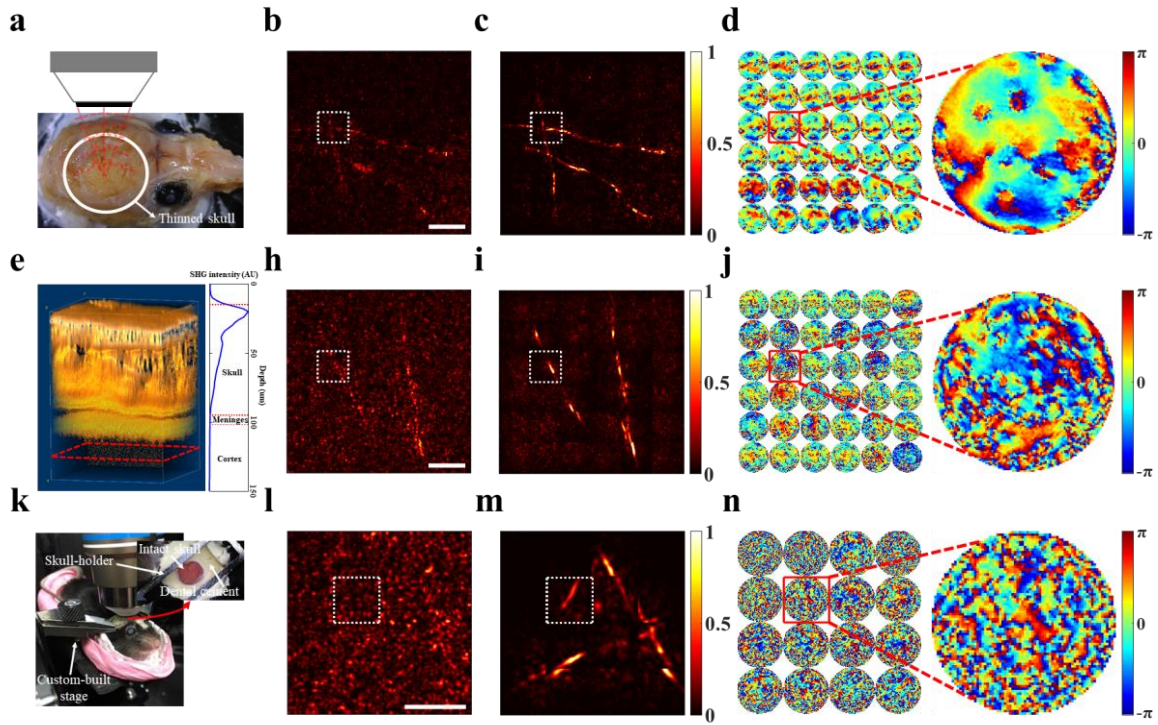

**Figure 3. Imaging of myelinated axons through a mouse skull. a-d,** Ex vivo imaging of mouse brain through a thinned skull of a 7-week-old mouse. The thickness of the skull was approximately 40-μm thickness. **a,** Ex vivo specimen of a fixed mouse head with a thinned skull. **b,** Conventional OCM image of myelinated fibers in cortex layer 1 at a depth of 70 μm from the upper surface of the thinned skull. Scale bar: 10 μm. **c,** Aberration-corrected LS-RMM image. The CLASS algorithm was individually applied to 6×6 subregions to correct local aberrations, and the corrected images were stitched together. The size of each subregion is 10×10 μm$^2$ including the overlapping area with the neighbors. **d,** Pupil phase maps retrieved for the subregions. The magnified map is the pupil aberration for the subregion indicated by the white dotted box in **b** and **c**. **e–h,** Ex vivo imaging through an intact skull of a 3-week-old mouse. The thickness of the skull was about 80μm. **e,** 3D reconstruction of SHG imaging of an intact skull. **h,** OCM image of myelinated fibers at 125 μm below the upper surface of the skull, which is marked by the red dotted box in **e**. Scale bar, 10 μm. **i** and **j**, LS-RMM image and retrieved pupil phase maps, respectively. The image was analyzed in the same way as in **c**. **k-n,** In vivo imaging of mouse brain through an intact skull of an 8-week-old mouse. The thickness of the skull was 125-150 μm. **k,** Experimental setup for in vivo imaging of a living mouse. **l,** OCM image of myelinated fibers at a depth of 200 μm from the upper surface of the skull. Scale bar, 10 μm. **m** and **n,** LS-RMM image and retrieved pupil phase maps, respectively. The image was analyzed by dividing the total view field of 30×30 μm$^2$ into 4×4 subregions. The size of each subregion is 10×10 μm$^2$ including overlapping margins.

**Figure 4**

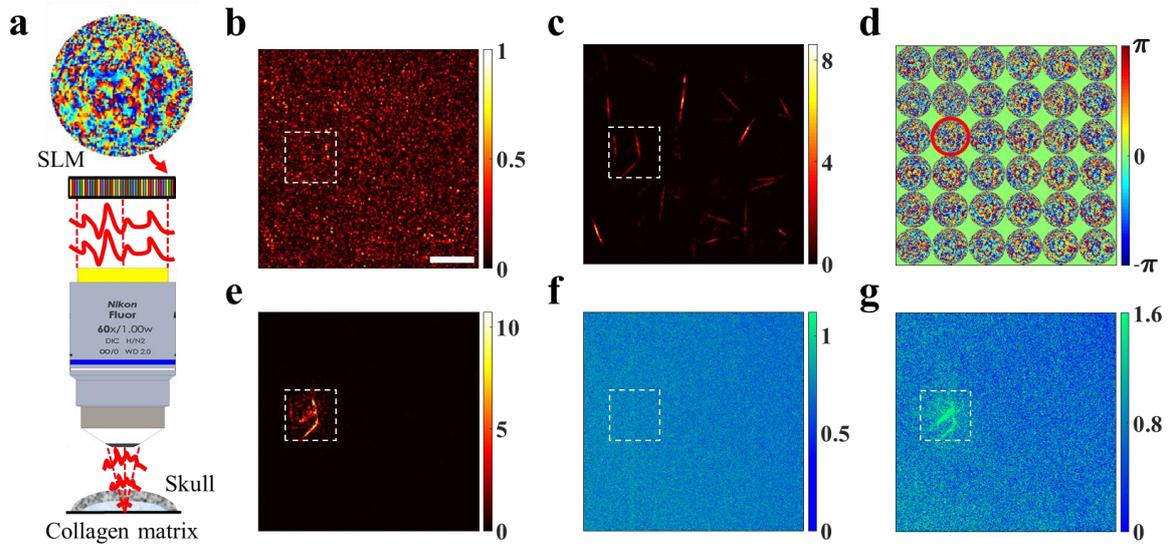

**Figure 4. SHG imaging of collagen fibers through an intact mouse skull. a,** Experimental schematic. Collagen gel matrix was placed under the ex vivo mouse skull from a 3-week-old mouse. The thickness of the skull was about 100 μm. **b,** Conventional OCM image without aberration correction. Scale bar, 10 μm. **c,** LS-RMM image stitched after applying aberration correction to each of 6×6 subregions. **d,** Pupil phase maps for the subregions. **e,** OCM image obtained after physically correcting the pupil aberration indicated by the red circle in **d** by using the SLM. **f** and **g,** SHG images before and after physically correcting the pupil aberration by using the SLM, respectively. The white dashed box in each figure indicates the subregion where the physical aberration correction was applied. Color bars in **b, c** and **e** indicate intensity normalized by the maximum intensity in **b**. Color bars in **d** and **e**, phase in radians.

**References**
1. Minsky, M. Microscopy Apparatus. *US Patent 3013467* **3013467**, 5 (1957).
2. Dunsby, C. & French, P. M. W. Techniques for depth-resolved imaging through turbid media including coherence-gated imaging. *J. Phys. D. Appl. Phys.* **36**, R207–R227 (2003).
3. Pawley, J. *Handbook of Biological Confocal Microscopy*. (Springer US, 2010).
4. Izatt, J. A., Swanson, E. A., Fujimoto, J. G., Hee, M. R. & Owen, G. M. Optical coherence microscopy in scattering media. *Opt. Lett.* **19**, 590 (2008).
5. Booth, M. J. Adaptive optical microscopy: the ongoing quest for a perfect image. *Light Sci. &Amp; Appl.* **3**, e165 (2014).
6. Tyson, R. K. *Principles of Adaptive Optics*. (CRC Press, 2015).
7. Girkin, J. M., Poland, S. & Wright, A. J. Adaptive optics for deeper imaging of biological samples. *Curr. Opin. Biotechnol.* **20**, 106–110 (2009).
8. Booth, M. J. Adaptive optics in microscopy. *Philos. Trans. R. Soc. A Math. Phys. Eng. Sci.* **365**, 2829–2843 (2007).
9. Park, J.-H., Kong, L., Zhou, Y. & Cui, M. Large-field-of-view imaging by multi-pupil adaptive optics. *Nat. Methods* **14**, 581–583 (2017).
10. Park, J.-H., Sun, W. & Cui, M. High-resolution in vivo imaging of mouse brain through the intact skull. *Proc. Natl. Acad. Sci.* **112**, 9236–9241 (2015).
11. Wang, C. *et al.* Multiplexed aberration measurement for deep tissue imaging in vivo. *Nat. Methods* **11**, 1037–1040 (2014).
12. Scrimgeour, J. & Curtis, J. E. Aberration correction in wide-field fluorescence microscopy by segmented-pupil image interferometry. *Opt. Express* **20**, 14534 (2012).
13. Ji, N., Milkie, D. E. & Betzig, E. Adaptive optics via pupil segmentation for high-resolution imaging in biological tissues. *Nat. Methods* **7**, 141 (2009).
14. Jian, Y. *et al.* Wavefront sensorless adaptive optics optical coherence tomography for in vivo retinal imaging in mice. *Biomed. Opt. Express* **5**, 547 (2014).
15. Wang, K. *et al.* Rapid adaptive optical recovery of optimal resolution over large volumes. *Nat. Methods* **11**, 625–628 (2014).
16. Aviles-Espinosa, R. *et al.* Measurement and correction of in vivo sample aberrations employing a nonlinear guide-star in two-photon excited fluorescence microscopy. *Biomed. Opt. Express* **2**, 3135 (2011).
17. Cha, J. W., Ballesta, J. & So, P. T. C. Shack-Hartmann wavefront-sensor-based adaptive optics system for multiphoton microscopy. *J. Biomed. Opt.* **15**, 046022 (2010).
18. Roorda, A. *et al.* Adaptive optics scanning laser ophthalmoscopy. *Opt. Express* **10**, 405 (2002).
19. Zheng, W. *et al.* Adaptive optics improves multiphoton super-resolution imaging. *Nat. Methods* **14**, 869–872 (2017).
20. Rueckel, M., Mack-Bucher, J. A. & Denk, W. Adaptive wavefront correction in two-photon microscopy using coherence-gated wavefront sensing. *Proc. Natl. Acad. Sci.* **103**, 17137–17142 (2006).
21. Adie, S. G., Graf, B. W., Ahmad, A., Carney, P. S. & Boppart, S. A. Computational adaptive optics for broadband optical interferometric tomography of biological tissue. *Proc. Natl. Acad. Sci.* **109**, 7175–7180 (2012).
22. Hillmann, D. *et al.* Aberration-free volumetric high-speed imaging of in vivo retina. *Sci. Rep.* **6**, 35209 (2016).
23. Kumar, A., Drexler, W. & Leitgeb, R. A. Subaperture correlation based digital adaptive optics for full field optical coherence tomography. *Opt. Express* **21**, 10850 (2013).
24. Papadopoulos, I. N., Jouhanneau, J. S., Poulet, J. F. A. & Judkewitz, B. Scattering compensation by focus scanning holographic aberration probing (F-SHARP). *Nat. Photonics* **11**, 116–123 (2017).
25. Liu, T.-L. *et al.* Observing the cell in its native state: Imaging subcellular dynamics in multicellular organisms. *Science (80-. ).* **360**, eaaq1392 (2018).
26. Pande, P., Liu, Y.-Z., South, F. A. & Boppart, S. A. Automated computational aberration correction method for broadband interferometric imaging techniques. *Opt. Lett.* **41**, 3324 (2016).

27. Kang, S. *et al.* High-resolution adaptive optical imaging within thick scattering media using closed-loop accumulation of single scattering. *Nat. Commun.* **8**, 2157 (2017).
28. Kim, M. *et al.* Label-free neuroimaging in vivo using synchronous angular scanning microscopy with single-scattering accumulation algorithm. *Nat. Commun.* **10**, 3152 (2019).
29. Choi, C., Song, K.-D., Kang, S., Park, J.-S. & Choi, W. Optical imaging featuring both long working distance and high spatial resolution by correcting the aberration of a large aperture lens. *Sci. Rep.* **8**, 9165 (2018).
30. Kang, S. *et al.* Imaging deep within a scattering medium using collective accumulation of single-scattered waves. *Nat. Photonics* **9**, 253–258 (2015).
31. Badon, A., Boccara, A. C., Lerosey, G., Fink, M. & Aubry, A. Multiple scattering limit in optical microscopy. *Opt. Express* **25**, 28914 (2017).
32. Badon, A. *et al.* Smart optical coherence tomography for ultra-deep imaging through highly scattering media. *Sci. Adv.* **2**, e1600370 (2016).
33. Tang, J., Germain, R. N. & Cui, M. Superpenetration optical microscopy by iterative multiphoton adaptive compensation technique. *Proc. Natl. Acad. Sci.* **109**, 8434–8439 (2012).
34. Wang, T. *et al.* Three-photon imaging of mouse brain structure and function through the intact skull. *Nat. Methods* **15**, 789–792 (2018).
35. Tehrani, K. F., Kner, P. & Mortensen, L. J. Characterization of wavefront errors in mouse cranial bone using second-harmonic generation. *J. Biomed. Opt.* **22**, 1–10 (2017).